\documentclass[amsmath,amssymb,reprint]{revtex4-1}
\usepackage[utf8]{inputenc}
\usepackage[T1]{fontenc}
\usepackage{mathptmx}
\usepackage{geometry}
\usepackage[colorlinks=true]{hyperref}
\usepackage{placeins}
\usepackage{booktabs}
\usepackage{color}
\usepackage{natbib}
\usepackage{latexsym,bm}
\usepackage{amsfonts}
\usepackage{graphicx}
\usepackage{longtable}
\begin{document}
\title{Temperature-dependent properties of liquid-vapour coexistence system with many-body dissipative particle dynamics with energy conservation}

\author{Kaixuan Zhang}
\affiliation{School of Aerospace Engineering and Applied Mechanics, Tongji University, Shanghai 200092, China}
\author{Jie Li}
\affiliation{School of Aerospace Engineering and Applied Mechanics, Tongji University, Shanghai 200092, China}

\author{Shuo Chen}
\email[Corresponding Authors:~]{schen\_tju@mail.tongji.edu.cn}
\affiliation{School of Aerospace Engineering and Applied Mechanics, Tongji University, Shanghai 200092, China}

\author{Yang Liu}
\affiliation{Department of Mechanical Engineering, The Hong Kong Polytechnic University, Hong Kong, China}

\begin{abstract}
The dynamic properties of fluid, including density, surface tension, diffusivity and viscosity, are temperature-dependent and can significantly influence the flow dynamics of mesoscopic non-isothermal systems. To capture the correct temperature-dependence of a fluid, a many-body dissipative particle dynamics model with energy conservation (mDPDe) is developed by combining the temperature-dependent coefficient of the conservative force and weighting terms of the dissipative and random forces. The momentum and thermal diffusivity, viscosity, and surface tension of liquid water at various temperatures ranging from 273 K to 373 K are used as examples for verifying the proposed model. Simulations of a periodic Poiseuille flow driven by body forces and heat sources are carried out to validate the diffusivity of the present model. Also, a steady case of heat conduction for reproducing the Fourier law is used to validate the thermal boundary conditions. By using this mDPDe simulations, the thermocapillary motion of liquid water nanodroplets on hydrophobic substrates with thermal gradients is investigated. The migration of the droplet is observed on flat substrates with gradient temperature. The velocity of the migration becomes larger for higher temperature difference, which is in agreement with the present theoretical analysis and DVDWT simulations. The results illustrate that the modified model can be used to study Marangoni effect on a nanodroplet and other heat and mass transfer problems with free interface. 
\end{abstract}


\maketitle

\section{introduction}
Dissipative particle dynamics is a particle-based model which was proposed by Hoogerbrugge and Koelman~\cite{1992Hoogerbrugge} and has been developed a suitable method for micro- and nano simulations. By combining the advantages of large timescale in lattice-gas automata (LGA), and the mesh-free property in molecular dynamics (MD), DPD is more efficient for simulating larger fluid system than MD. Moreover, DPD is defined as a coarse-grained model, which means every DPD particle represents a cluster of atoms/molecules and the computation is greatly reduced. Due to the larger spacial and time scale, DPD becomes a flexible method for investigate simple fluids~\cite{2006_Chen} and complex fluid problems such as polymer~\cite{2009_Li} and DNA suspensions~\cite{2006_Fan}, red blood cells dynamics~\cite{2013_Peng, Ye2014Dissipative}, and biofluids~\cite{Xu2011Dissipative}. 

The interaction between DPD particles determines the basic properties of fluid system, which involving conservative force, dissipative force and random force. The conservative force is a soft repulsive force, which makes contribution to the fluid compressibility. The dissipative force generates friction between DPD particles, which can describe the viscosity of the system. The random force makes up the defect due to coarse-graining treatment by introducing a stochastic force on each DPD particle and meanwhile, these two forces satisfy the fluctuation-dissipation theorem and act as a thermostat to control the system at a constant temperature~\cite{1997Groot}. Since there is no attractive force between DPD particles, the equation of state does not match van der Waals curve and can not simulate vapour-liquid coexistence system. By introducing attractive force term and adding local density in the repulsive term, the modified conservative force produces a $many-body$ interaction with $long~range$ attraction and $short~range$ repulsion~\cite{2003_Warren}. The pressure of the system becomes a cubic relationship with density, which can generate a free interface. This many-body dissipative particle dynamics method has been widely used to simulate droplet dynamics on substrates, multiphase flow in complex geometry and phase transition problems~\cite{2011_Arienti, 2012_Li_Dissipative, Mo2017Hypernetted, Zhao2018Rheology, 2018_Pan, Yuxiang2019Droplet, 2019_kai}. 

Both the classic DPD and its $many-body$ version, i.e. mDPD, can only describe isothermal system due to the thermostat in the models. To simulate the process of heat transfer, an extra temperature quantity is introduced on each DPD particle and generate energy-conserved dissipative particle dynamics (eDPD)~\cite{Espa_ol_1997}, which has been adopted to investigate heat conduction~\cite{2007_Qiao}, natural convection~\cite{Abu2010Natural}, and solidification~\cite{2016Modeling}. Ripoll $et~al.$~\cite{1998_Ripoll_Dissipative} confirmed that the heat conduction of eDPD system obeys Fourier law by simulating the 1D heat conduction problem. He $et~al.$~\cite{He2008Self} adopted eDPD to investigated the process of heat conduction between nano-fluid and nano-materials. Abu-Nada $et~al.$~\cite{Abu2010Natural} considered Neumann and Dirichlet boundary conditions and simulate the 2D heat conduction problem by using eDPD method. They also modified the parameters in eDPD model to study natural convection and then validated their results by comparing with CFD simulations. Li $et~al.$~\cite{Li2014Energy} proposed analytical formula for determining the mesoscopic heat friction and validated the prediction by reproducing the experimental data for Prandtl number of liquid water at various temperatures. Johansson $et~al.$~\cite{2016Modeling} adopted eDPD to simulate phase transfer problem, and they captured the process of solidification which is agreement with theoretical analysis. They also suggested that a modified model is needed to reproduce the release of the latent heat during the process of the transition from liquid to solid phase.

To extend the isothermal mDPD equations to modeling heat transport in non-isothermal multiphase fluid systems, many-body dissipative particle dynamics with energy conservation (mDPDe) was developed by combining mDPD with eDPD. Yamada $et~al.$~\cite{Yamada2016Dissipative} simulated the process of heat conduction between droplets and solid substrates by using mDPDe. The droplet density at different temperature seems to be constant in their simulations, which does not match that of most liquids in the real world. Wang $et~al.$~\cite{2020_Wang_zhang} adopted mDPDe to investigate ice crystal nucleation leading to droplet freezing on flat substrates. They captured the nucleation and shape deformation process of a water droplet during freezing which are in agreement with the experimental data. Their work illustrated the modified mDPDe can be used to study freezing of water droplet. However, the corresponding temperature-dependent properties of the system at various temperature are still not clear in the numerical model, which are also important to illustrate the validity of the model. 

The objective of the present work is to propose a model for capturing the correct temperature-dependent properties of fluid system with free interface, such as Marangoni effect on a droplet. Specifically, liquid water is used as an example for verifying the mDPDe model we propose. The density, surface tension, diffusivity and viscosity of liquid water as well as its Schmidt and Prandtl numbers in the range of 273K to 373K are reproduced with the present mDPDe model. The relationship between the surface tension and the expressions of the conservative force have been analyzed. The temperature-dependent density and surface tension are obtained, which can be important for investigating fluid problems with complex mass transfer. The relation of the weight function with temperature proposed by Li $et~al.$ is adopted~\cite{Li2014Energy}. Results for the diffusivity and viscosity as well as Schmidt number at various temperatures are presented and compared with the available experimental data of liquid water. Furthermore, a fitting formula for correcting the mesoscopic heat friction is obtained, which reproduces more accurate Prandtl number of liquid water at various temperatures. This proposed model is not limited to liquid water, and it can be readily extended to other fluids for modelling the correct dependence of thermal hydrodynamic properties in terms of temperature. 

The paper is organized as follows. In Section 2 we describe details of mDPDe formulations and parameters, including analytical formulas for determining the surface tension and the mesoscopic heat friction. Section 3 presents our validation of the mDPDe model, and the performance of the present mDPDe model in reproducing correctly the temperature-dependent properties consistent with the experimental data. Furthermore, we simulate the thermalcapillary motion of a droplet on a hydrophobic substrate with temperature gradients and compare the velocity with theoretical analysis and DVDWT simulations. Finally, we conclude with a brief summary in Section 4. 

\section{Method}
Many-body dissipative particle dynamics model with energy conservation(mDPDe) is developed based on many-body dissipative particle dynamics (mDPD) and energy conservation dissipative particle dynamics(eDPD). Each mDPDe particle interacts with other particles through distance- and velocity-dependent forces and energies within a certain cutoff radii~\cite{1992Hoogerbrugge, Espa_ol_1997, J1997Dissipative}. The momentum and energy transfer between mDPDe particles obey the momentum and energy conservation, respectively. The evolution of the particle motion is controlled by Newton's second law:
\begin{equation}m_{i} \frac{d \vec{v}_{i}}{d t}=\sum_{j \neq i}\left(\vec{f}_{i j}^{C}+\vec{f}_{i j}^{D}+\vec{f}_{i j}^{R}\right)+\vec{f}_{e x t}
\end{equation}
where $\vec{f}_{e x t}$ is the external force. $\vec{f}_{i j}^{C}$, $\vec{f}_{i j}^{D}$, and $\vec{f}_{i j}^{R}$ denote the conservative, dissipative and random forces among particles, respectively. Their specific expressions are as follows:
\begin{equation}\vec{f}_{i j}^{C}=A_{i j} \omega^{C}\left(r_{i j}\right) \vec{e}_{i j}+B_{i j}\left(\tilde{\rho}_{i}+\tilde{\rho}_{j}\right) \omega^{d}\left(r_{i j}\right) \vec{e}_{i j}\end{equation}

\begin{equation}\vec{f}_{i j}^{D}=-\gamma_{i j} \omega^{D}\left(r_{i j}\right)\left(\vec{e}_{i j} \cdot \vec{v}_{i j}\right) \vec{e}_{i j}\end{equation}

\begin{equation}\vec{f}_{i j}^{R}=\sigma_{i j} \omega^{R}\left(r_{i j}\right) \zeta_{i j} \Delta t^{-1 / 2} \vec{e}_{i j}\end{equation}
where $A_{ij}$ and $B_{ij}$ are the amplitudes of attractive and repulsive forces, respectively. $\omega^{C}\left(r_{i j}\right) = \left(1 - r_{ij}/r_{C}\right)$ and $\omega^{d}\left(r_{ij}\right) = \left(1 - r_{ij}/r_{d}\right)$ represent the distance-dependent weight functions which vanish if the distance is larger than the corresponding cut-off radii, $r_C$ and $r_d$, respectively. $\tilde{\rho_i}$ and $\tilde{\rho_j}$ are the weighted local density functions of particles $i$ and $j$, respectively, which are described as $\tilde{\rho}_{i}=\sum_{j \neq i} \omega^{\rho}\left(r_{i j}\right)$. And in the description, the weight function of the local density is expressed as $\omega^{\rho}\left(r_{ij}\right)={105}/{16 \pi r_{d}^{3}} \cdot \left(1+{3 r_{i j}}/{r_{d}}\right)\left(1-{r_{ij}}/{r_{d}}\right)^{3}$, and $\int_{0}^{\infty} d^{3} \mathbf{r} \omega^{\rho}\left(r_{ij}\right) = 1$  as the weight function is normalized by the factor ${105}/{16 \pi r_{d}^{3}}$. Another choice is $\omega^{\rho}\left(r_{i j}\right)={15}/{2 \pi r_{d}^{3}} \cdot \left(1-{r_{i j}}/{r_{d}}\right)^{2}$, which can provide smaller density for the system. In this work, we adopt the former one. $\zeta_{i j} = \zeta_{j i}$ is a random number with zero mean value and a variance of unity, which keeps the momentum of the interacting pair of particles to be conserved. The dissipative coefficient $\gamma_{ij}$ and random coefficient $\sigma_{ij}$ satisfy the fluctuation-dissipation theorem, which requires:
\begin{equation}\sigma_{i j}^{2}=\frac{4 \gamma_{i j} k_{B} T_{i} T_{j}}{\left(T_{i}+T_{j}\right)}, \quad \omega^{R}\left(r_{i j}\right)=\left[\omega^{D}\left(r_{i j}\right)\right]^{1 / 2}\end{equation}
where $k_B$ and $T$ are Boltzmann constant and system equilibrium temperature, respectively. $\omega^{D}\left(r_{i j}\right)$ and $\omega^{R}\left(r_{i j}\right)$ are the weight functions for the dissipative and random forces, respectively, in which $\omega^{D}\left(r_{i j}\right) = \left(1-r_{ij}/r_{C}\right)^{s_v}$ for $r_{ij} < r_{C}$. $s_v$ is related with the viscosity of fluid and here, we adopt the modified expression proposed by Li $et~al.$~\cite{Li2014Energy} to obtain temperature-dependent viscosity. 

As the method of MDPDe satisfies the energy conservation equation, the heat transfer between particles is realized by the exchange of internal energy, which is an additional property and expressed as follows~\cite{Espa_ol_1997}:
\begin{equation}C_{v} \frac{d T_{i}}{d t}=\sum_{j \neq i}\left(q_{i j}^{V_{h}}+q_{i j}^{C_{h}}+q_{i j}^{R_{h}}\right)\end{equation}   
where the heat flux, $q_i$, is the sum of collision-caused heat flux $q_{ij}^C$ , mechanical energy-caused viscous heating $q_{ij}^V$ and thermal fluctuation-caused heat flow $q_{ij}^R$  within the truncation radius. $C_V$ is the heat capacity at constant volume, which is normalized by $k_B$. The specific expressions are given by
\begin{equation}q_{i j}^{V_{h}}=\frac{1}{2 C_{v}}\left[\omega^{D}\left(r_{i j}\right)\left\{\gamma_{i j}\left(\vec{e}_{i j} \cdot \vec{v}_{i j}\right)^{2}-\frac{\sigma_{i j}^{2}}{m_{i}}\right\}-\sigma_{i j} \omega^{R}\left(r_{i j}\right)\left(\vec{e}_{i j} \cdot \vec{v}_{i j}\right) \zeta_{i j}\right]\end{equation}
\begin{equation}q_{i j}^{C_{h}}=\kappa_{i j} \omega^{CT}\left(r_{i j}\right)\left(\frac{1}{T_{i}}-\frac{1}{T_{j}}\right)\end{equation}
\begin{equation}q_{i j}^{R_{h}}=\alpha_{i j} \omega^{RT}\left(r_{i j}\right) \zeta_{i j}^{e} \Delta t^{-1 / 2}\end{equation}               
where $k_{ij}$ and $\alpha_{ij}$ represents the strengths of conductive and random heat fluxes respectively, which are given by
\begin{equation}\kappa_{i j}=k_{o} k_{B} T_{e q}^{2}\left(\frac{\epsilon_{i}+\epsilon_{j}}{2 k_{B} T_{e q}}\right)^{n_{k}}\end{equation}
\begin{equation}\alpha_{i j}=\sqrt{2 k_{B} \kappa_{i j}}\end{equation}
where $\epsilon_{i}$ is the interal energy. $k_{o}$ is a positive constant which determines the thermal conductivity of mDPDe particles. $T_{eq}$ is the equilibrium temperature of the system and $n_{\kappa}$ is the constant which was chosen as 2 in the previous work~\cite{2007_Qiao, Abu2010Natural}. $\kappa$ is a constant related to the thermal conductivity of mDPDe particals, and $\omega^{CT} \left({r_{ij}}\right)$ and $\omega^{RT}\left(r_{ij}\right)$ are the weight functions, respectively and their relation can be described as $\omega^{CT} \left({r_{ij}}\right) = \left[\omega^{RT}\left(r_{ij}\right)\right]^{2}$ and $\omega^{CT} \left({r_{ij}}\right) = \left(1-r_{ij}/r_{CT}\right)^{s_T}$. $s_T$ is assigned as 2 in the previous paper~\cite{Li2014Energy}. $\zeta_{i j}^{e} = - \zeta_{j i}^{e}$ is the random number having similar properties as $\zeta_{i j}$, which also keeps the energy of the interacting pair of particles to be conserved~\cite{Espa_ol_1997}.

\section{Results and Discussion}
\subsection{Validation of Fourier law}
\begin{figure}[hbpt!]
\includegraphics[width = 0.3\textwidth]{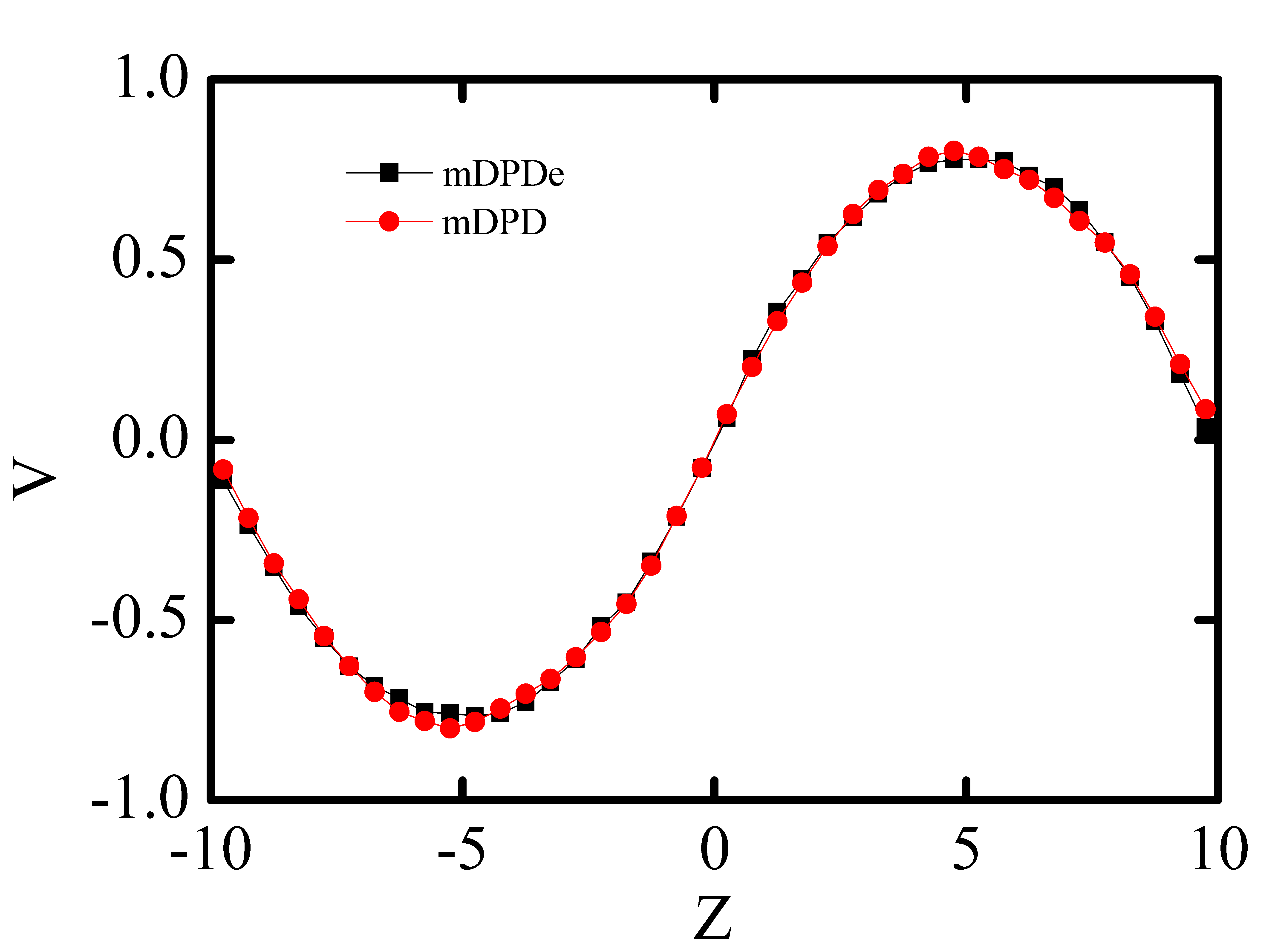}
\caption{The velocity profile of the periodic Poiseuille flow along the $z$-direction with the attractive parameter $A = -40$, the repulsive parameter $B = 25$, the temperature $k_BT = 1.0$, the dissipative parameter $\gamma = 8.0$, and the external force $F = 0.02$.}
\label{fig_PPF}
\end{figure}

To validate the hydrodynamic property of the present non-isothermal mDPDe model, we simulate periodic Poiseuille flow with a simple mDPDe fluid and compare the results with isothermal mDPD model. Two equal force in reverse direction $\left(F = 0.02\right)$ are applied on each mDPDe particle to drive the flow. The mDPDe parameters for the periodic Poiseuille flow are listed in the caption of Fig.~\ref{fig_PPF}. The computational domain is divided into 50 bins along the $z$-direction. The equilibrium velocity profiles is shown in Fig.~\ref{fig_PPF}, which are obtained by averaging enough sampled data.  The results are in agreement with that of mDPD model, which illustrates that we can use the modified mDPDe model to obtain the same hydrodynamic property with mDPD.
\begin{figure}[hbpt!]
\includegraphics[width = 0.25\textwidth]{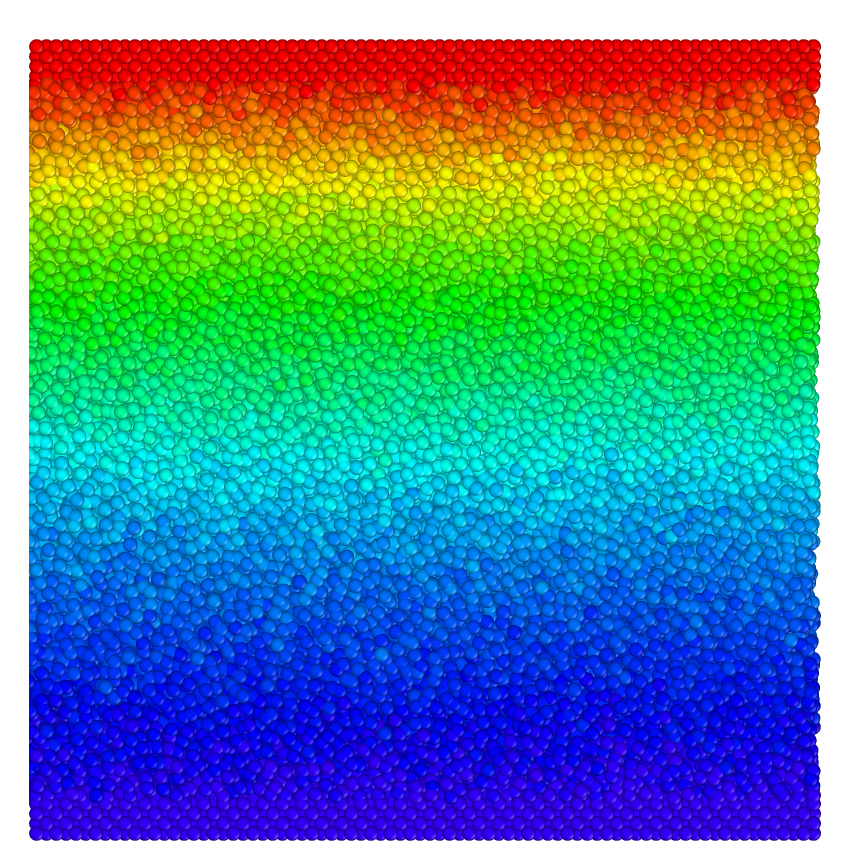}
\caption{Schematic of the geometry for the steady-state heat conduction between a cold wall $T_C$ (blue) and a hot wall $T_H$ (red).}
\label{fig_hc}
\end{figure}

\begin{figure*}
\centering
\includegraphics[width = 0.85\textwidth]{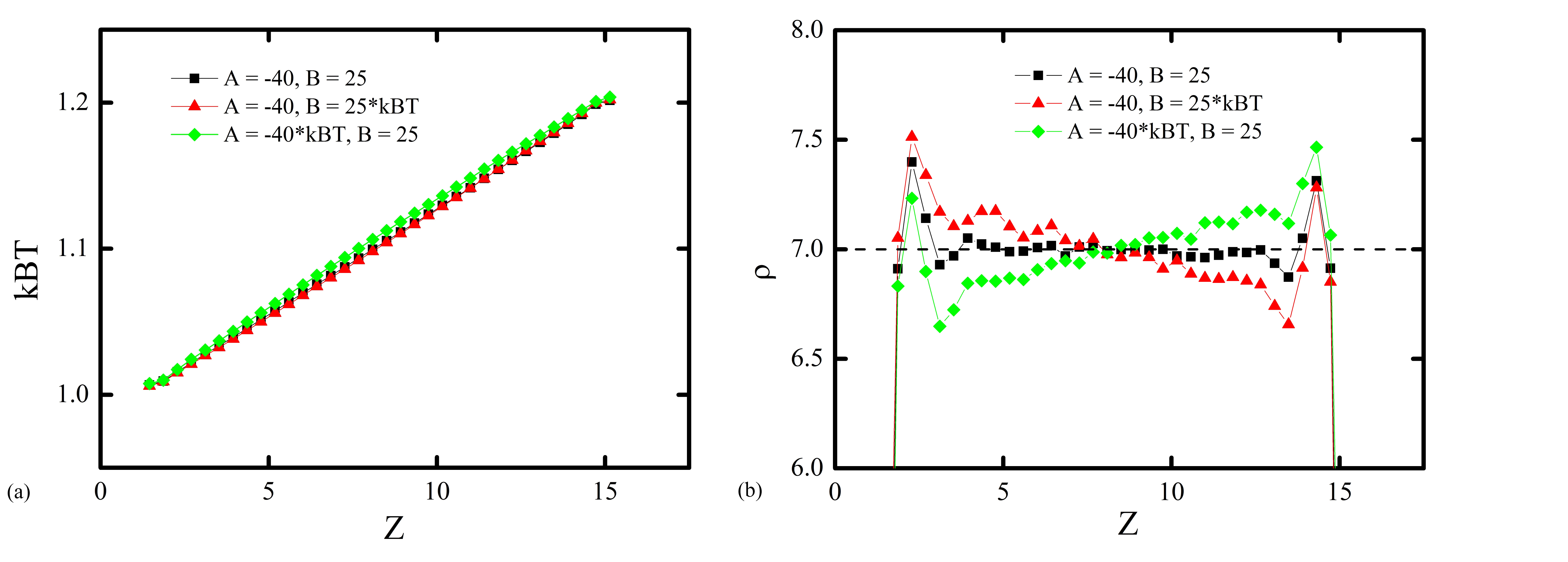}
\caption{Temperature profiles in the heat conduction between a hot wall $\left(T_H = 1.2\right)$ and a cold wall $\left(T_C = 1.0\right)$ with different conservative parameters $A$ and $B$. The simulations use 18500 mDPDe particles in a computational domain $40.0~\times~4.0~\times~17.0$ in mDPDe units.}
\label{fig_kBT}
\end{figure*}
The next test case for validation is heat conduction between a cold wall and a hot wall. The schematic of the geometry is shown in Fig.~\ref{fig_hc}, in which the stationary fluid is confined between a hot wall of $T_H = 1.2$ and a cold wall of $T_C = 1.0$. We test different parameters, including constant attractive parameter $A = -40$ and repulsive parameter $B = 25$; constant $A = -40$ and temperature-dependent $B = 25*k_BT$; and temperature-dependent $A = -40*k_BT$ and constant $B = 25$. Fig.~\ref{fig_kBT}~(a) shows the temperature profile of the fluid from all these tests has a linear spatial distribution for the steady-state heat conduction, which obeys the Fourier law. However, as shown in Fig~\ref{fig_kBT}~(b), the density in these systems change differently with the increasing temperature in the system. Referring to the real density profile of water, the density decreases with increasing temperature in most range from 273~K to 373~K, a temperature-dependent repulsive parameter $B = 25 * k_BT$ and a constant value of attractive parameter $A = -40$ is more proper for simulating the heat conduction between liquid water and substrates.

\subsection{Temperature-dependent properties}
The above two cases validate the modified mDPDe model. For an isothermal fluid system, the present model captures the correct behaviours of fluid as mDPD does. For a non-isothermal fluid system, the present model can reproduce Fourier law in heat conduction between a hot wall and a cold wall. Our next objective is to construct a model for capturing the correct dynamic properties of common fluids in non-isothermal systems.
\begin{figure}[hbpt!]
\includegraphics[width = 0.4\textwidth]{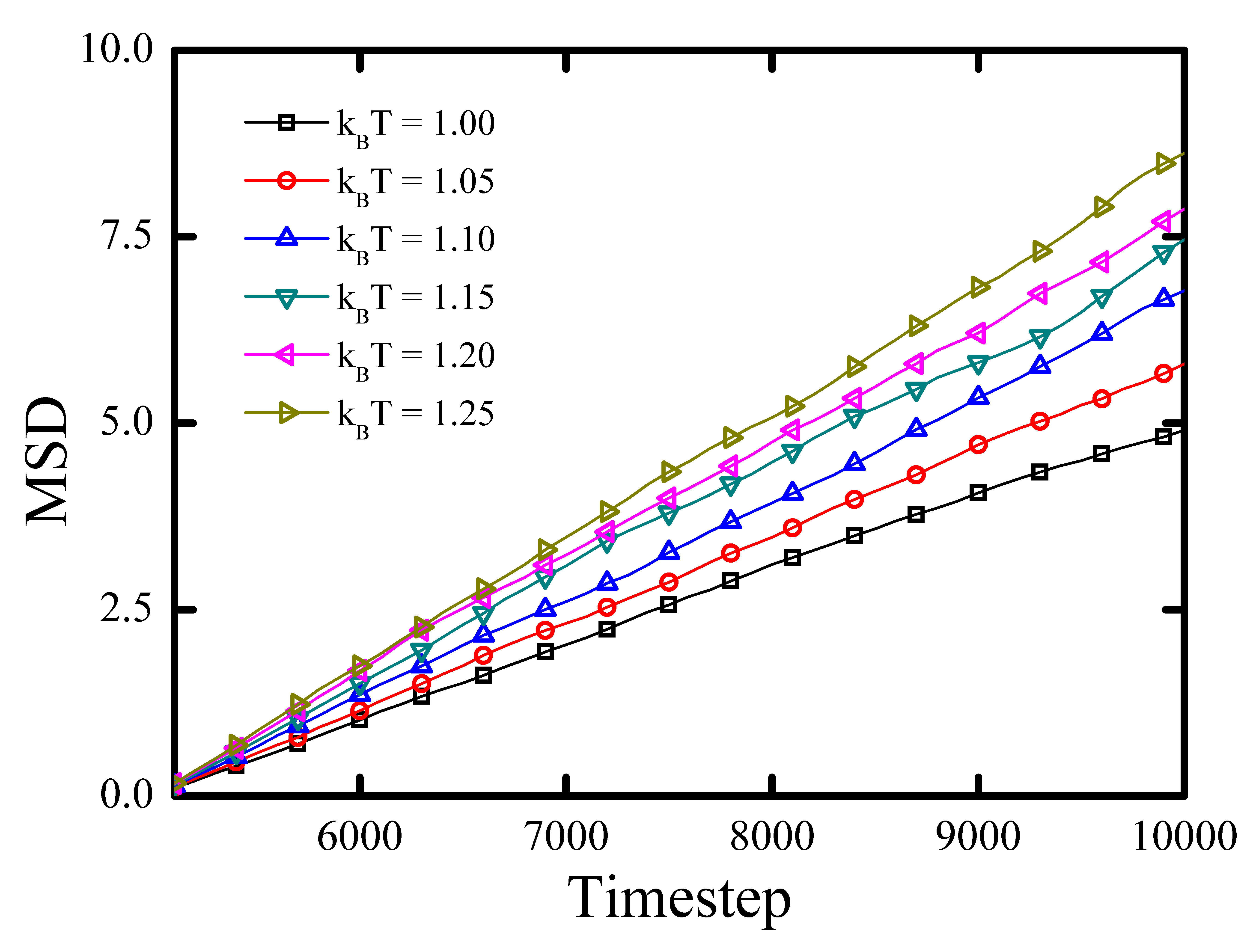}
\caption{Mean-square displacement (scaled by 6.0) of mDPDe particles for different temperatures. The slopes of the lines are the self-diffusivity of mDPDe fluid.}
\label{fig_MSD_timestep}
\end{figure}

The liquid water is employed as an example fluid in the present study. We test the self-diffusivity, kinematic viscosity, surface tension, density, and thermal diffusivity at various temperature, which are all output properties instead of input parameters. And their exact values can be only obtained by simulating mDPDe system. The compressibility of liquid system can be described as $\kappa_{c}^{-1}=(\partial p / \partial \rho)_{T} / k_{B} T$. As the pressure of the mDPD fluid system can be approximated as 
$p=\rho k_{B} T+\alpha A \rho^{2}+2 \alpha B r_{d}^{4}\left(\rho^{3}-c \rho^{2}+d\right)$, in which $\alpha=0.101 \pm 0.001$, $c=4.16 \pm 0.02$, and $d=18 \pm 1$~\cite{2003_Warren}. The compressibility of the fluid system can be derived as $\kappa_{c}^{-1}= 1 + 1/k_BT\left[2\alpha A \rho + 2\alpha B r_d^4\left(3\rho^2-2c\right)\right]$. On the other hand, the dimensionless coefficient corresponding to the compressibility of liquid system is $\kappa_{c}^{-1}={[L]^{3}}/{\rho k_{B} T \beta_{T}}$~\cite{1997Groot}. For liquid water at 300~K, the thermal term is $k_{B} T=4.142 \times 10^{-21} \mathrm{kgm}^{2} \mathrm{s}^{-2}$ and $\beta_{T}=4.503 \times 10^{-10} \mathrm{ms}^{2} / \mathrm{kg}$. For a mDPDe system with the density $\rho = 6.8$, $A = -40$, and $B = 25$, the corresponding scaling length is $[L] \approx 1.18~nm$. The dimensionless coefficient is approximated as 130, which is larger than that from the pressure~(76.3). Thus, the compressibility of the fluid system could be underestimated, as we adopt $A = -40$ and $B = 25 * k_BT$ to test the basic properties including density $\rho$, surface tension $\sigma$, and kinematic viscosity $\nu$ and scaling the units by comparing these parameters with those of water at different temperatures. The underestimation can be accepted in some degree as the error from the density and pressure measurement could be magnified by the coefficients in the relationship between the compressibility and the parameters of the mDPDe system. As for the diffusivity of the system, We determine the self-diffusivity of the mDPDe system by the mean-square displacement 
\begin{equation}D=\lim _{t \rightarrow \infty} \frac{1}{6 t}\left\langle|\mathbf{r}(t)-\mathbf{r}(0)|^{2}\right\rangle\end{equation}
where $|\mathbf{r}(t)-\mathbf{r}(0)|^{2}$ is the mean-square displacement (MSD). Fig.~\ref{fig_MSD_timestep} shows the MSD of the system at $T = 1.0-1.25$. The MSD has been scaled by 6.0 thus the slope of the line is the self-diffusivity for each case. It can be found in this Figure that the diffusivity increases when the temperature increases. 
\begin{figure}[hbpt!]
\includegraphics[width = 0.4\textwidth]{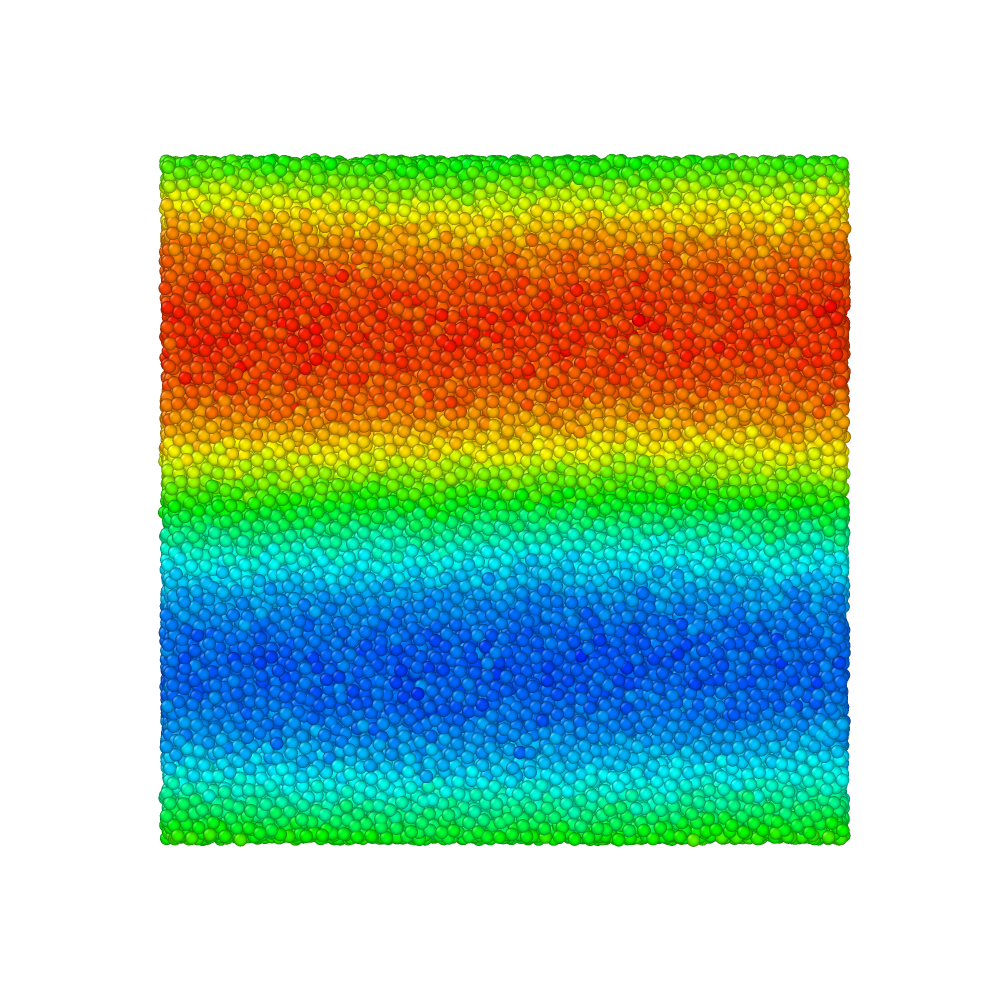}
\caption{Schematic of $heat~conduction~analog~$of periodic Poiseuille flow~\cite{Li2014Energy}}
\label{fig_HC_PPF_1}
\end{figure}
\begin{figure}[hbpt!]
\includegraphics[width = 0.4\textwidth]{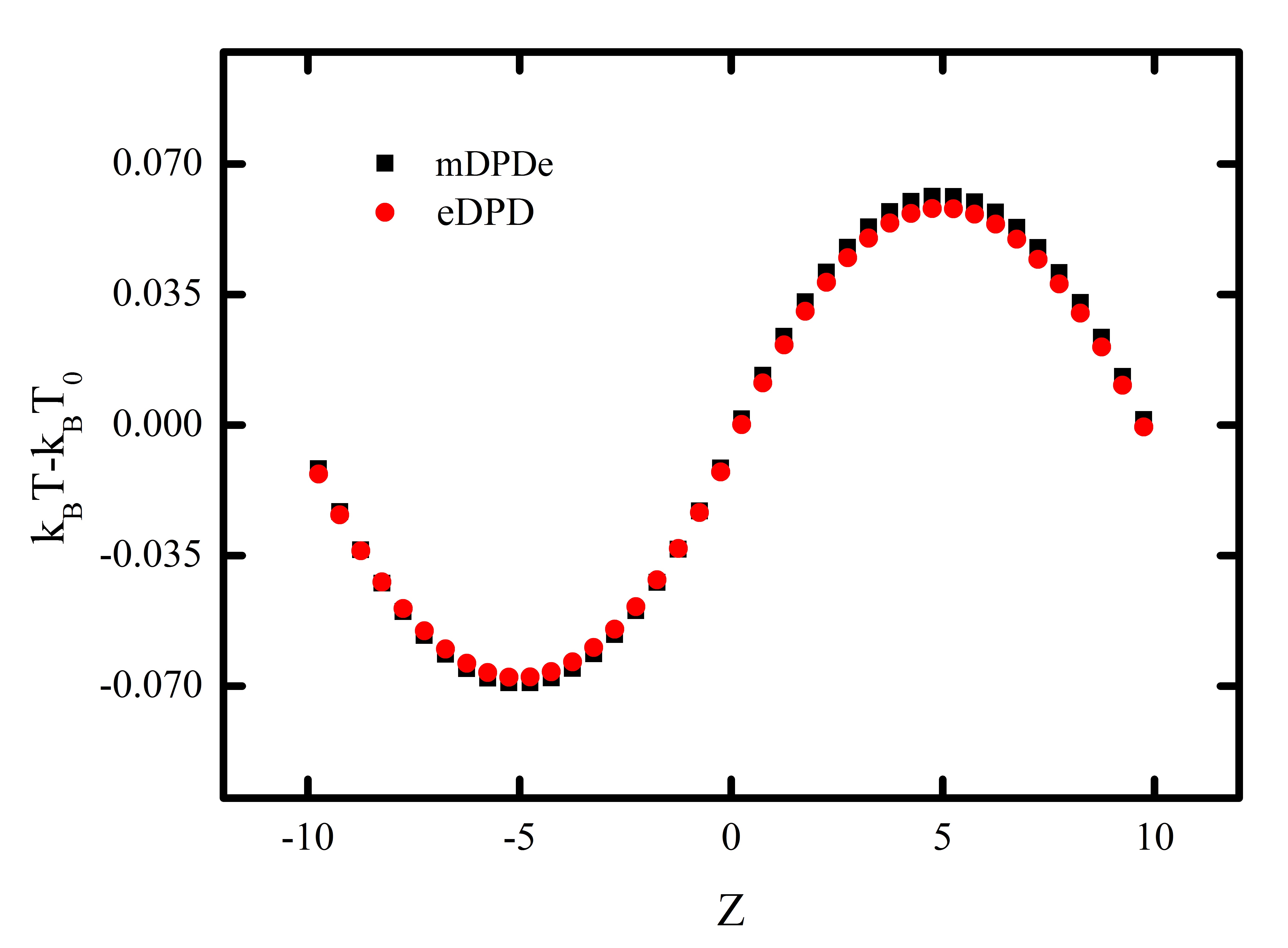}
\caption{Comparison of the temperature profiles obtained using the $heat~conduction~analog$ of periodic Poiseuille flow for different temperatures between the results of mDPDe and eDPD simuations.}
\label{fig_HC_PPF_2}
\end{figure}   
\begin{table}[hbpt!]
\begin{tabular}{@{}lllll@{}}
\toprule
 $T_i$ (K) &  kBT & $\sigma$ & $\nu$ & $\rho$ \\ \midrule
 273 & 0.91 & 9.93  & 7.71 & 7.00 \\ \midrule
 283 & 0.94335 & 9.63  & 5.49 & 6.92 \\ \midrule
 293 & 0.97667 & 9.34  & 4.04 & 6.81 \\ \midrule
 313 & 1.0433 & 8.76 & 2.91 & 6.62 \\ \midrule
 333 & 1.11 & 8.09 & 1.89 & 6.43 \\ \midrule
 353 & 1.17667 & 7.33 & 1.41 & 6.25 \\ \midrule
 373 & 1.2433 & 6.93 & 1.01 & 6.08 \\ \bottomrule
\end{tabular}
\caption{The basic properties of mDPDe model at various temperatures.}
\label{tab:my-table}
\end{table}
\begin{figure}[hbpt!]
\includegraphics[width = 0.4\textwidth]{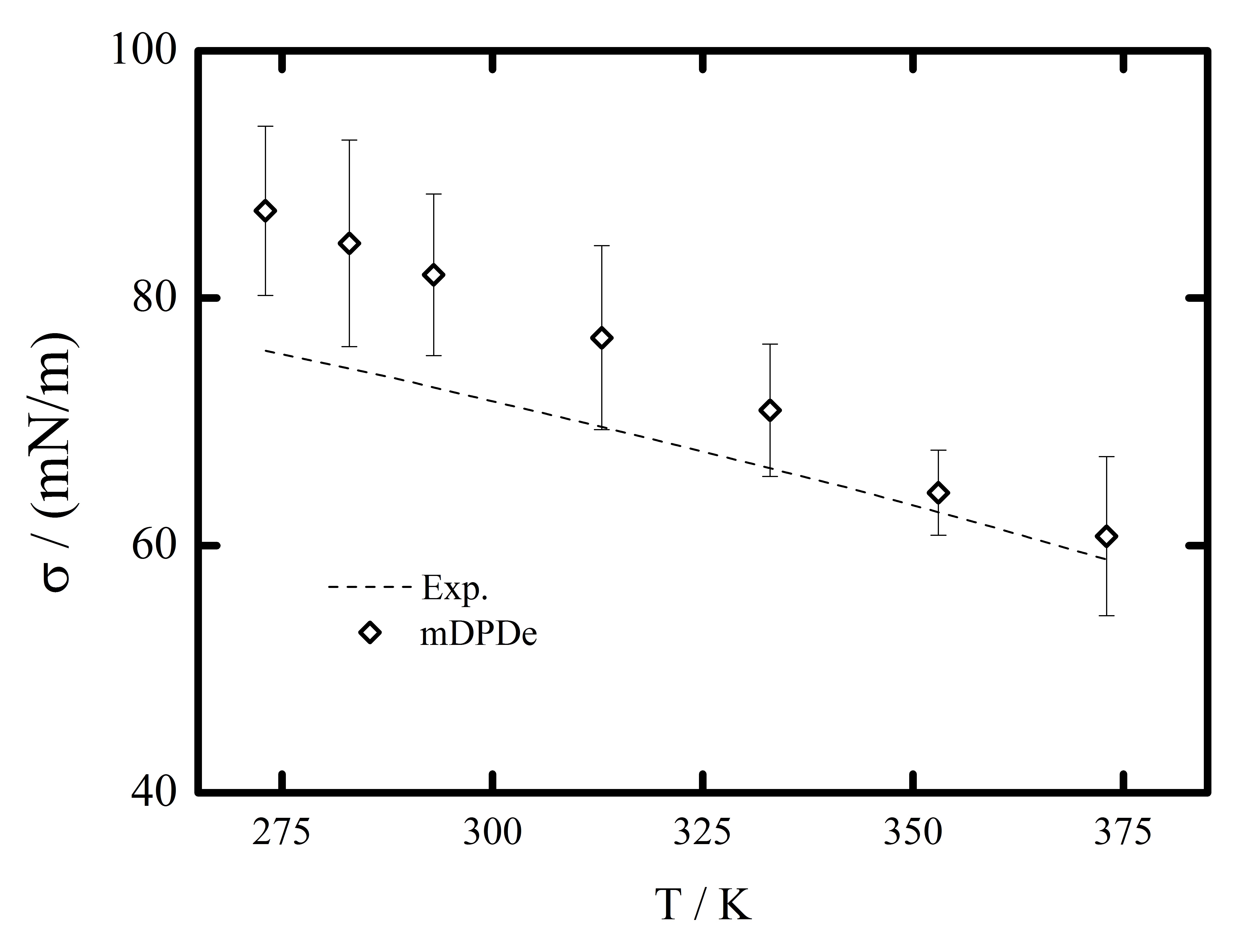}
\caption{Comparison of the temperature-dependent surface tension $\sigma$ for the temperature ranging from 273~K to 373~K between the experimental data and the results of mDPDe simulations.}
\label{fig_sigma}
\end{figure}

\begin{figure*}
\centering
\includegraphics[width = 0.8\textwidth]{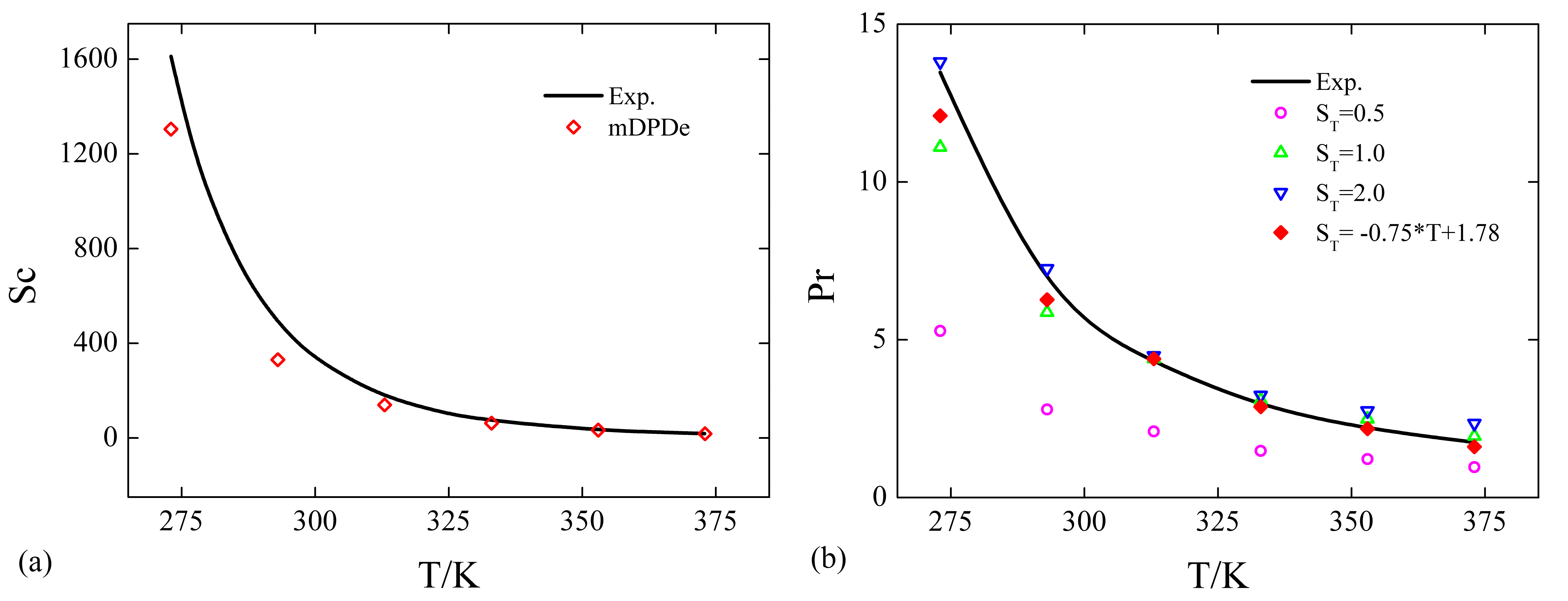}
\caption{Comparison of the temperature-dependent Schmidt number (a) and Prandtl number (b) for the temperature ranging from 273~K to 373~K between the experimental data of liquid water and the results of mDPDe simulations.}
\label{fig_Sc_Pr}
\end{figure*}

The temperature unit was characterized as $T_R = 300K$. The mass, time, and length units were characterized as~\cite{2011_Arienti, 2019_kai}~$M_{D P D}=L_{D P D}^{3} {d^{*}}/{d}$, $T_{D P D}=\left(M_{D P D} {\sigma}/{\sigma^{*}}\right)^{1/2}$, and ${L_{D P D}^{2}}/{T_{D P D}}={v^{*}}/{v}$. The basic properties of water are marked as density $\rho$, surface tension $\sigma$, and viscosity $\nu$, while the corresponding parameters in mDPDe system are those with asterisk. For every group of properties including density, viscosity, and surface tension, we can obtain corresponding length [L]$\left(T_i\right)$, mass [M]$\left(T_i\right)$, and time [T]$\left(T_i\right)$ units according to the systems of equations mentioned above. Based on the test results which are shown in Table~\ref{tab:my-table}, a series of these basic units can be obtained. Due to all the length units [L]$\left(T_i\right)$ fluctuates within one order at around $10^{-9}~m$, we adopt the average value of the series results derived from the table to describe the length unit of the system at the temperature of 273~K to 373~K, as well as the time and mass units. Thus, the length, time, and mass units of the system are respectively $[\bar{L}] = 1.18 \times 10^{-9}~m$, $[\bar{T}] = 5.58 \times 10 ^{-12}~s$, and $[\bar{M}] = 2.73 \times 10^{-25}~kg$. The kinematic viscosity of the mDPDe system is computed from the periodic Poiseuille flow method. The velocity profile obtained for the periodic Poiseuille has been shown in Fig.~\ref{fig_PPF}. The kinematic viscosity can be determined by fitting the velocity profile with the analytical solution $u(z)=g_{x} z(d-|z|) / 2 \nu$ in which $\nu$ is the kinematic viscosity. $g_{x}=0.02$ is the body-force applied on mDPDe particles and $d = 10.0$ is the half-length of the computational domain in $z$-direction. Based on the previous scaling, we can obtain the suraface tension from mDPDe simulations, as shown in Fig.~\ref{fig_sigma}, which is in good agreement with that in experimental data. Moreover, From the simulations we find the corresponding Schmidt number are also in good agreement with that of the real water, as shown in Fig.~\ref{fig_Sc_Pr}~(a). Overall, the proposed mDPDe model reproduces the correct temperature-dependent dynamic properties including the diffusivity and the kinematic viscosity as well as Schmidt number comparable to the experimental data of liquid water for the temperatures ranging from 273~K to 373~K. The relative errors of Schmidt number are generally less than 10$\%$ of experimental data.

For the thermal conductivity in the liquid system, we adopt the $heat~conduction~analog$ of periodic Poiseuille flow method which was proposed by Li $et~al.$~\cite{Li2014Energy} to obtain the thermal diffusivity of mDPDe fluid ranging from 273~K to 373~K. The schametic of the geometry and the temperature profile are shown in Fig.~\ref{fig_HC_PPF_1} and~\ref{fig_HC_PPF_2}. And the result is represented by Prandtl number $Pr = \nu/\lambda$, which is the ration of momentum diffusivity $\nu$ to thermal diffusivity $\lambda$. Fig.~\ref{fig_Sc_Pr}~(b) shown the comparison of the temperature-dependent Prandtl number between the results of mDPDe simulations and the experimental data of liquid water. It can be observed that for different the parameter $s_T$, the error or difference between the numerical results and experimental data is different. In particular, for $s_T = 2.0$, which is adopted as the default value, the error at high temperature is larger than 15~$\%$. While for $s_T = 1.0$, the values of Prandtl number at low temperature is lower than the experimental data over than 18$~\%$. Thus, we test a series of values of $s_T$ at various temperatures and then obtain an empirical formula to describe accurate Prandtl number of mDPDe model, i.e. $s_T = -0.75*T+1.78$. It can be observed that the Prandtl number of mDPDe fluid are consistent with the experimental Prandtl numbers. The relative errors is less than 10~$\%$ of experimental $Pr$ for the entire temperature range of 273K to 373K. 

\subsection{Thermal-capillary motion of a droplet}

\begin{figure}[hbpt!]
\includegraphics[width = 0.4\textwidth]{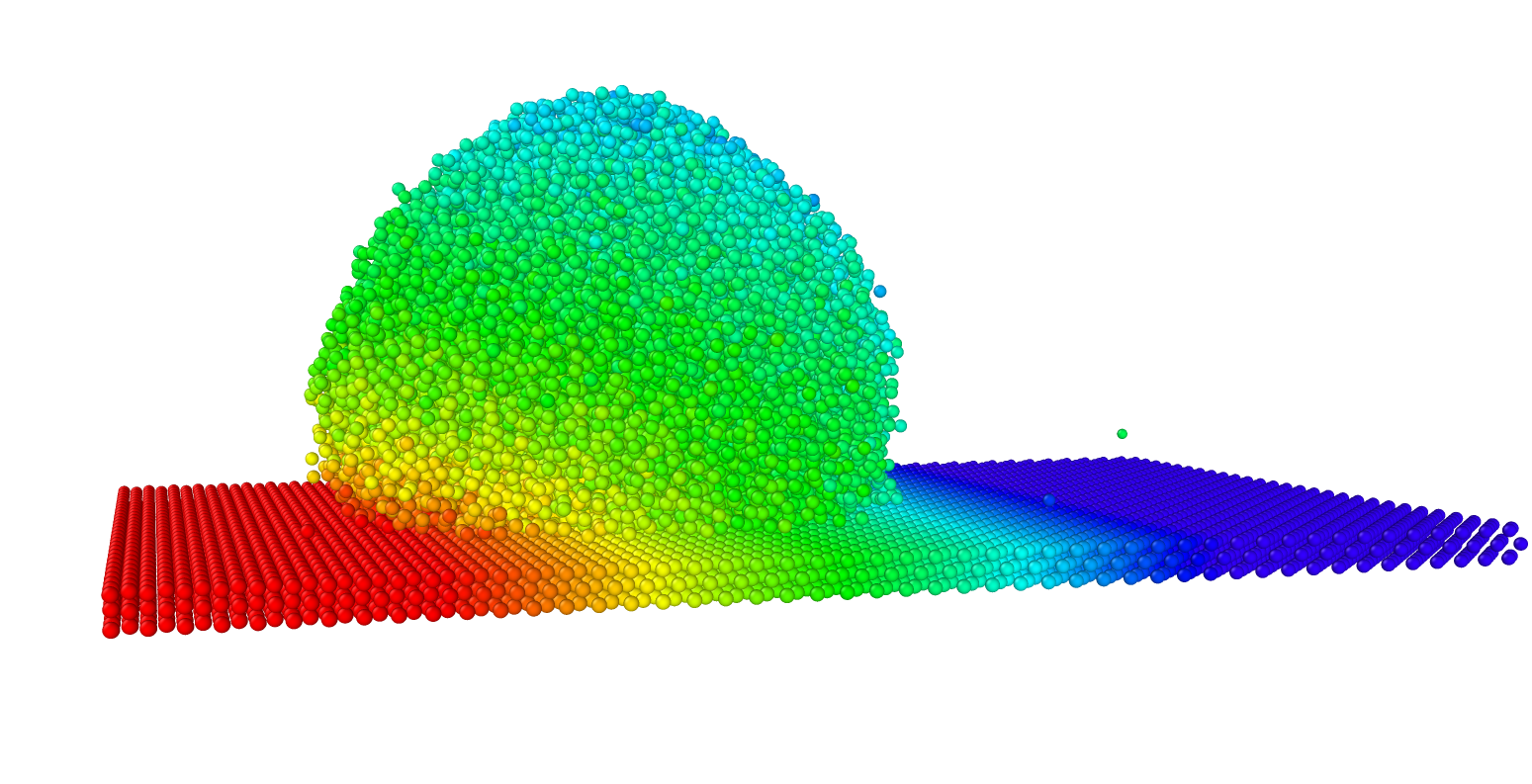}
\caption{Schematic of droplet motion induced by thermal gradient on a hydrophobic substrate. The temperature of the hot end (red) is $T = 360~K$, and the cold end (blue) $T = 300~K$. The length of the substrate is around $70~nm$, which has a temperature gradient of $0.857~K/nm$.}
\label{fig_DT}
\end{figure}

Based on the modified mDPDe, we investigate thermalcapillary motion of a droplet on a hydrophobic substrate with a temperature gradient, as shown in Fig.~\ref{fig_DT}. This phenomenon has been investigated by other theoretical, experimental and numerical methods in the previous work~\cite{1989_Motions, 2012_Xu}, which suggested that the moving velocity is dependent on the the temperature difference between the ends of the droplets along the temperature gradient. The interaction between liquid-liquid particles in the conservative force is $A = -40, B = 25*k_BT$ and the interaction between solid-liquid particles are $A_{sl} = -15$ and $B_{sl} = 12.5$, for which the static contact angle is around 120 degree. 
\begin{figure}[hbpt!]
\includegraphics[width = 0.4\textwidth]{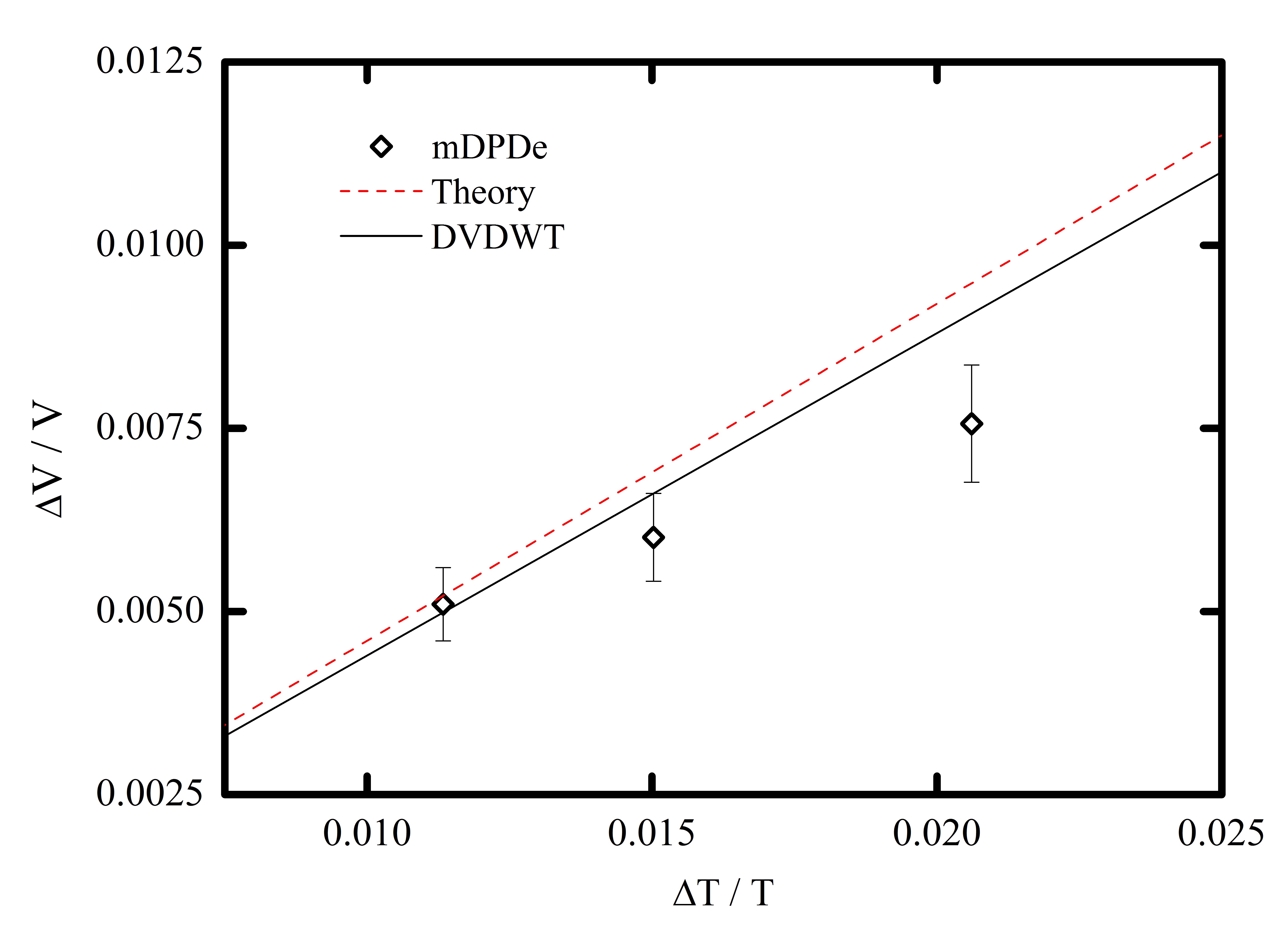}
\caption{Comparison with theoretial analysis and DVDWT simulations~\cite{2012_Xu}. The velocity $V = 400~m/s$ and the temperature $T = 600~K$, which have been noted in the Ref.~\cite{2012_Xu}.}
\label{fig_velocity}
\end{figure}

In the simulations, we firstly set a droplet on a hydrophobic substrate with constant temperature $k_BT = 1.0$. After the droplet get its equilibrium state on the substate, i.e. forming a spherical cap (around 10000 timesteps), the temperature gradient is applied on the substrate and then the droplet can spontaneous move from the hot to the cold area along the substrate due to the Marangoni effect (around 90000 timesteps). The timestep is set as 0.005 and then the total time of Marangoni effect is around 450 in mDPDe units. We test three different cases with various sizes of droplets on the same substrate. Based on the scaling as mentioned above, the sizes of the water droplets in the simulations are around $8.3~nm$ including 8061 mDPDe particles, $9.5~nm$ with 14000 mDPDe particles, and $11~nm$ with 22408 mDPDe particles, respectively. The length of the substrate is around $70~nm$ along the $x$-direction and $45~nm$ along the $y$-direction and $1.2~nm$ along the $z$-direction which includes 14400 mDPDe particles. The temperature of the susbtrate is from $360~K$ to $300~K$ along the $x$-direcion. And the gradient of the temperature is around $0.857~K/nm$. Bounce-back boundary condition is used in this model which has been used in the previous works~\cite{Yuxiang2019Droplet}. We measure the distance of droplet motion during the process of Marangoni effect and then calculate the velocity of the droplets which shows the velocity of droplets increases with larger temperature difference along the temperature gradients. It is noting that the process of Marangoni effect can be affected by the random number selection in the mDPDe simulations, which was also illustrated in the previous work~\citep{2011_Arienti}. By comparing with the theoretical analysis and numerical results with DVDWT method~\cite{2012_Xu}, as shown in Fig.~\ref{fig_velocity}, this modified mDPDe model can effectively capture the thermalcapillary motion of nanodroplets on hydrophobic substrates with a temperature gradient. In the comparison, the velocity $V = 400 m/s$ and the temperature $T = 600 K$ for water, which are used to scale the variables~\cite{2012_Xu}.

\section{Conclusions}
A modified many-body dissipative particle dynamics with energy conservation (mDPDe) model for reproducing correctly the temperature-dependent properties including density, surface tension, Schmidt and Prandtl number has been proposed, which can be adopted to investigate the droplet motion on a substrate induced by Marangoni effect. The relationships between liquid-vapour surface tension of mDPDe fluid and the parameters in the conservative force are analyzed. Combining with the weighting terms of the dissipative and random forces, the temperature-dependent self-diffusivity and thermal diffusivity are obtained. The liquid-vapour surface tension, viscosity, momentum and thermal diffusivity of liquid water at various temperature ranging from 273~K to 373~K were used as a benchmark for verifying the model. The results show that the present model reproduces the correct temperature-dependent properties including liquid-vapour surface tension, Schmidt number, and Prandtl number consistent with the available experimental data of liquid water. Moreover, an corrected formula for obtaining more accurate Prandtl number at high temperature has been obtained from mDPDe simulations.

Hence, this work proposes a non-isothermal mDPD model with energy conservation (mDPDe) which is able to reproduce the correct temperature-dependent liquid-vapour surface tension, Schmidt number and Prandtl number. Furthermore, this mDPDe model is used to simulate the thermalcapillary motion of a droplet on a hydrophobic substrate with a temperature gradient. The results show that the velocity of droplets increases with larger temperature difference between the advancing and receding triple-phase contact area, which is in good agreement with the theoretical analysis and DVDWT simulations~\cite{2012_Xu}. Although we test our model with liquid water, the method proposed in the present work for modelling the correct temperature-dependent properties is not limited to water only and it can be readily extended to other fluids.

\section{Acknowledgement}
This work was supported by the National Natural Science Foundations of China~(Grant No.\ 11872283). K. Zhang would like to acknowledge Dr. Jiayi Zhao and Mr. Damin Cao for their helpful discussions and Dr. Chensen Lin for his technological help. 
\bibliographystyle{unsrt}
\bibliography{references}

\end{document}